\documentstyle[psfig]{mn_plus_bib}

\def\Mpc{\,\hbox{Mpc}}

\def\Msun{\hbox{M}_{\odot}}
\def\kms{\,\hbox{km}\,\hbox{s}^{-1}}
\def\kpc{\,\hbox{kpc}}
\def\H0{H_0=100 \, h \, {\rm kms^{-1}Mpc^{-1}}}
\def\Zsol{Z_{\odot}}
\def\SN{S_N}

\def\etal{{et al.\thinspace}}
\def\eg{{e.g.\thinspace}}

\def\ie{{i.e.\thinspace}}

\def\B{{\emph{B}}}

\newif\ifAMStwofonts

\begin{document}

\title[
Modelling Centaurus A
]
{
Simultaneous Modelling of the Stellar Halo and Globular Cluster System of NGC 5128
}
\author[
M. A. Beasley et al.
]
{
Michael. A. Beasley$^{1}$\thanks{email:
mbeasley@astro.swin.edu.au}, William E.~Harris$^{2,4}$,
Gretchen L.~H.~Harris$^{3,4}$,
\newauthor
Duncan A. Forbes$^{1}$\\
$^1$ Astrophysics \& Supercomputing, Swinburne University,
Hawthorn VIC 3122, Australia\\
$^2$ Department of Physics and Astronomy, McMaster University, 
Hamilton ON L8S 4M1, Canada\\
$^3$ Department of Physics, University of Waterloo,
Waterloo ON N2L 3G1, Canada \\
$^4$ Visiting Fellow, Research School of Astronomy \& Astrophysics,
Australian National University, Weston ACT 2611, Australia \\
}

\pagerange{\pageref{firstpage}--\pageref{lastpage}}
\def\LaTeX{L\kern-.36em\raise.3ex\hbox{a}\kern-.15em
    T\kern-.1667em\lower.7ex\hbox{E}\kern-.125emX}

\newtheorem{theorem}{Theorem}[section]

\label{firstpage}

\maketitle

\begin{abstract}
An important test for models of galaxy formation
lies in the metallicity distribution functions (MDFs) of 
spheroid stars and their globular clusters (GCs).
We have compared the MDFs obtained from
spectroscopy of the GCs and the star-by-star photometry of the old halo 
red giants in the nearby elliptical galaxy NGC~5128, with the predictions 
of a $\Lambda$CDM semi-analytic galaxy formation model.
We have selected model ellipticals comparable in
luminosity and environment to NGC~5128, and reconstructed 
their MDFs by summing the total star formation occurring
over all their progenitors. A direct comparison between models
and data shows that the MDFs are qualitatively similar, both 
have stellar components which are predominantly metal-rich 
($\sim$0.8$\Zsol$), with a small fraction of metal-poor 
stars extending down to 0.002$\Zsol$. 
The model MDFs show only small variations between systems, 
whether they constitute brightest cluster galaxies or 
low luminosity group ellipticals.
Our comparison also reveals that these model MDFs harbour 
a greater fraction of stars at $Z>\Zsol$ than the observations, 
producing generally more metal-rich (by $\sim$0.1 dex) MDFs.
One possibility is that the outer-bulge observations 
are missing some of the highest-metallicity stars in this galaxy.
We find good agreement between the model and observed GC MDFs, 
provided that the metal-poor GC formation is halted early
($z\sim$5) in the model.
Under this proviso, both the models and data are bimodal with peaks at
0.1$\Zsol$ and $\Zsol$, and cover similar metallicity ranges.
This broad agreement for the stars and GCs 
suggests that the bulk of the stellar population
in NGC~5128 may have been built up in a hierarchical fashion,
involving both quiescent and merger-induced star formation. 
The predicted existence of age structure amongst the 
metal-rich GCs needs to be tested against high-quality
data for this galaxy. 
\end{abstract}

\begin{keywords}
galaxies: interactions -- galaxies: 
elliptical -- galaxies: evolution -- galaxies: individual: NGC 5128 -- 
galaxies: star clusters
\end{keywords}

\section{Introduction}

The task of understanding the star formation histories of 
elliptical galaxies is one of the most 
challenging areas of galaxy formation, 
and has generated a wide and steadily growing literature.  
For giant elliptical (gE) galaxies,
the formation history may be particularly complex 
because these systems are likely to involve
some (if not all) of the major formation processes that 
are currently debated, including singular or multi-phase
collapse (\eg \citeANP{Silk77} 1977; \citeANP{Arimoto87} 1987; 
\citeANP{Forbes97} 1997), the hierarchical merging of gaseous 
fragments in the early universe
(\eg \citeANP{Kauffmann93} 1993; \citeANP{Dubinski98} 1998;
\citeANP{Pearce99} 1999; \citeANP{Cole00} 2000; 
\citeANP{Somerville01} 2001), the major merging disc galaxies 
(\eg \citeANP{Schweizer87} 1987; \citeANP{Ashman92} 1992; 
\citeANP{Barnes92} 1992; \citeANP{Shioya98} 1998; \citeANP{Hibbard99} 1999;
\citeANP{Naab01} 2001; \citeANP{Bekki02} 2002) and ongoing
dissipationless accretion of satellites
(e.g. \citeANP{Cote98} 1998; \citeANP{Cote02} 2002).

Theoretical analyses employing sophisticated N-body simulations 
which include chemical evolution (\eg \citeANP{Shioya98} 1998; 
\citeANP{Steinmetz99} 1999; \citeANP{Koda00} 2000;
\citeANP{Kawata01} 2001; \citeANP{Lia02} 2002)
and semi-analytic treatments (\eg \citeANP{Cole94} 1994, 2000; 
\citeANP{Kauffmann96a} 1996; \citeANP{Somerville99} 1999; 
\citeANP{Thomas99} 1999) are
rapidly adding to our ability to understand this early 
sequence of events.  

However, observational constraints are badly needed 
to sort out the relative importance of these
possibilities. In this paper, we explore the ability of one 
hierarchical merging code to reproduce the metallicity
distributions of the recently obtained star-by-star
halo photometry and GC system of the nearby elliptical 
galaxy NGC~5128. 

At high-redshifts, there is some observational 
evidence that the bulk of the stars in at least 
a subset of galaxies may be in place as early as 
$z > 5$ (\eg\ \citeANP{Dunlop96} 1996; \citeANP{Franx97} 1997; 
\citeANP{Dey98} 1998; \citeANP{Nolan01} 2001;
\citeANP{Waddington02} 2002).  These studies
have been generally used to argue for single-collapse, high-redshift
models of gE galaxy formation. Other discussions 
suggest a somewhat longer timescale
and later epochs for much of the star formation
(\eg \citeANP{Zepf97} 1997;
\citeANP{Kauffmann98a} 1998; \citeANP{Barger99} 1999; 
\citeANP{Menanteau01} 2001 -- but also see \eg\ \citeANP{Jimenez99} 1999
for an different interpretation of the \citeANP{Zepf97} 1997 results).

On the observational side, an opportunity to test these 
models is through the metallicity distribution function 
(MDF) of the oldest stars in ellipticals.  
Indirect estimates of the MDF can be made 
through measures of the spectral features of their 
integrated bulge light (\eg \citeANP{Gonzalez93} 1993; 
\citeANP{Davies93} 1993; \citeANP{Harald98} 1999; 
\citeANP{Trager98} 1998; \citeANP{Terlevich02} 2002).  
However, two considerably more direct ways to study the 
MDF are now emerging as useful tools in the low redshift regime:
star-by-star photometry of the old-halo red giant stars; and 
metallicity measurements of the old-halo 
globular clusters (GCs) within the galaxies.  
Despite observational biases and shortcomings (see below),
these kinds of direct object-by-object construction of the MDF
provide a level of new information that cannot be obtained in 
any other way.

Neither of these latter approaches is immune from 
the interpretive difficulties associated with all such studies:
for example, metallicity measurements of either the old red-giant 
stars or GCs are uncertain to the degree that we 
know the particular mixture of ages present in the system, or
even whether a well defined age-metallicity relation (AMR) exists.
And for the GCs, other issues arise such as the efficiency
of cluster formation, which might in principle vary with metallicity 
or even the epoch of formation.

Recently, \citeANP{Beasley02} (2002) have applied a
semi-analytic galaxy formation code \cite{Cole00} 
to synthesise model MDFs for globular cluster systems (GCSs),
and have compared these with real GCSs in early-type galaxies.  
Detailed discussion of the elements of the model
can be found in these two papers.  In the model, a large 
galaxy is progressively assembled from a population of highly 
gaseous proto-galactic fragments (in the Cole et al. model, 
these are modelled as rotationally supported discs, 
and in this study we use the terminology of proto-galactic discs (PGDs) 
used by \citeANP{Beasley02} 2002).
The most important feature for our purposes is that   
the formation process, during which time star formation 
goes on simultaneously with the hierarchical merging 
of these PGDs, gives rise to two main 
subpopulations of objects:

\noindent {\it (i)} a population of stars and 
metal-poor GCs that are built within the PGDs
before they merge into the body of the larger central
galaxy (termed the ``quiescent mode'' of star
formation); and

\noindent {\it (ii)} a predominantly metal-rich population
built in successive star-bursts 
generated during major mergers (``burst mode''). 
This burst mode can happen any time that a PGD 
(in the fiducial model defined as a 
minimum of 30\% of the central
galaxy mass) merges with the central galaxy, and so can 
continue until quite late epochs.  

The quiescent mode of star formation generates
a distribution over age and metallicity that is fairly smooth and
continuous (star formation in this mode begins at $z \sim$ 12
in the fiducial model, and continues at low levels until
late epochs).
The burst mode also produces stars over a 
significant fraction of a Hubble time, although the onset of
merging begins after the first PGDs have formed 
from cooling gas.
In this mode, the stellar populations formed 
are more metal-rich due to their formation from enriched
gas, and are in the mean younger. 
Because these mergers are stochastically driven, the age
and metallicity distributions formed in bursts 
can differ widely from one galaxy to another.  

It is natural to associate the bimodal form of the 
GC MDF seen frequently for gE galaxies
with the two subpopulations described 
above (\eg \citeANP{Zepf93} 1993; \citeANP{Forbes97} 1997; 
\citeANP{Larsen01} 2001).
Typically, more or less equal numbers of GCs fall within each
mode, and the two modes have metallicity peaks consistently near
[Fe/H] $\simeq -1.5$ and $\simeq -0.5$ roughly independent of galaxy
luminosity.  In this interpretation, the ``blue'' GCs are 
seen as belonging to the earlier, quiescent mode while 
the ``red'', more metal-rich ones arise from the burst mode.

\begin{figure*}
\centerline{\psfig{figure=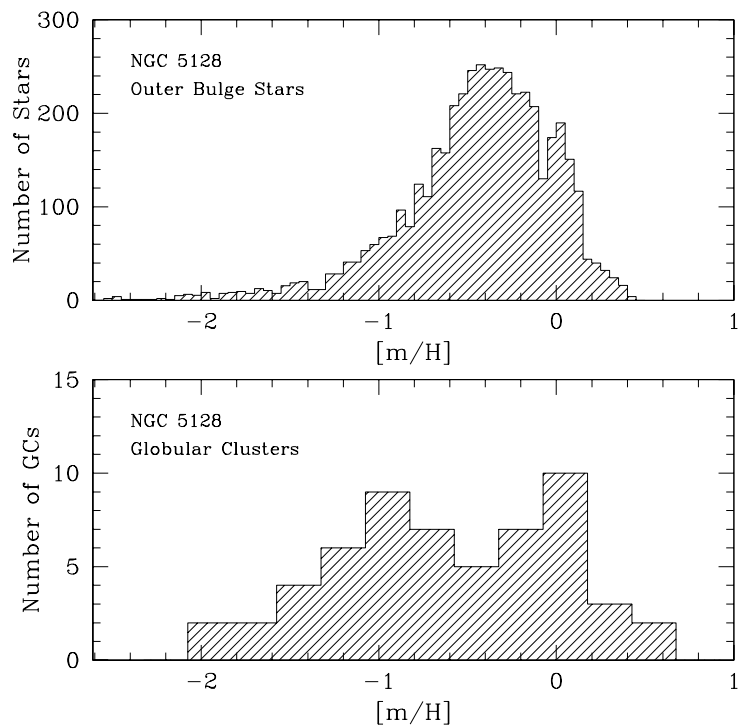,width=4.0in,height=4.0in}}
\caption{\label{fig1} 
{\sl:  Upper panel:} Observed [m/H] distribution for $\sim 17,000$ field
stars in NGC 5128 at a location 8 kpc SW of the galaxy centre,
from Harris \& Harris (2002).  This sample of stars is completely
dominated by an old red-giant population and has relatively high
mean metallicity, with very few stars more metal-poor than
[m/H] $\simeq$ --0.8.
{\sl Lower panel:}  Observed [m/H] distribution for GCs
in the halo of NGC 5128, from the spectroscopic sample
of Held et al.~(2002).  }
\end{figure*}  

However, \citeANP{Beasley02} (2002)
find that to predict the right numbers and metallicities of
the GCs in each mode, it is necessary to impose two extra 
numerical factors:  first, the {\sl formation times} 
of the blue GCs formed in the quiescent mode
must be truncated at an early epoch $z_{trunc}$  (otherwise they
would continue forming to later and later times at higher and 
higher metallicities, and their MDF would then end up 
overlapping substantially with the red GCs).
Second, the relative {\sl formation efficiency}, defined
as the mass ratio of GCs to all stars, 
$\epsilon=M_{GC}/M_{\star}$, needs to be different for the
two modes (otherwise the total numbers of 
GCs in each mode would be widely different).
For the well studied prototype case of NGC 4472,
\citeANP{Beasley02} (2002) find
best-fitting values $z_{trunc}\simeq5$, $\epsilon_R=0.007$,
and $\epsilon_B=0.002$.  
As we show later (Section~\ref{Bursts}) the numerical value of
$\epsilon_R$ is sensitive to a least one parameter in the
semi-analytic model, and should not be taken as an 'absolute value'.
The underlying physical reasons for enforcing the truncation
at $z\sim5$ is not yet understood, but within the parameters of the 
semi-analytic model, it is necessary to impose this condition
to maintain the observed numbers of GCs and 
the form of the MDF.

\begin{table*}
\begin{center}
\caption{Some basic properties of NGC~5128 and our selected model galaxies.}
\label{table1}
\begin{tabular}{llllllllll}
\hline 
Quantity 	& NGC~5128 	& Galaxy \#1 & \#2 & \#3 & \#4 & \#5 & \#6 & \#7 & \#8\\
\hline
M$_B$ 		& --21.0$^{a}$ 	&--22.63&--21.49&--21.28&--21.13&--21.06&--20.97&--20.61&--20.32\\
$\sigma (\kms)$ & 114$^{b}$ 	& 1460 & 368 & 286 & 330 & 244 & 288 & 247 & 228\\
$\mathcal{M/L}_B$ & 3.9$^{c}$ 	& 3.0 & 3.2 & 4.3 & 5.5 & 3.2 & 6.6 & 5.1 & 4.0\\
$(B-V)_0$ & 0.88$^{a}$ 		& 0.89	& 0.92 & 0.89 & 0.91 & 0.91 & 0.91 & 0.89 & 0.89\\
$(V-K)_0$ & 2.79$^{d}\dag$	& 2.93	& 3.02 & 2.91 & 2.97 & 2.97 & 3.00 & 2.94 & 2.93\\
M$_{\rm HI}$($\times10^{9}\Msun$)& 0.83$\pm$0.25 $^{e}$ & 3.1 & 2.1 & 1.0 & 1.3 & 0.49 & 1.5 & 0.065 & 1.3\\ 
$\SN$ 	     & 2.6$\pm$0.6$^{f}$& 5.8 & 4.1 & 3.8 & 2.8 & 3.7 & 2.5 & 2.0 & 2.7 \\
$p$(KS)$^{g}$	     &1.0&0.35&0.63&0.66&0.80&0.78&0.63 &0.78&0.35\\
\hline
\end{tabular}
\end{center}
$^{a}$ B-band absolute magnitude, integrated $B-V$
colour (\citeANP{RC3} 1991), 
 $^{b}$ Cen A group velocity dispersion (\citeANP{vandenBergh00} 2000),
$^{c}$ B-band mass-to-light ratio (\citeANP{Hui95} 1995), 
$^{d}\dag$ integrated $V-K$ colour (\citeANP{Pahre99} 1999),
calculated with unmatched apertures.
$^{e}$ Neutral gas mass (\citeANP{Richter94} 1994).
$^{f}$ GC Specific frequency (\citeANP{Harris91} 1991). 
$^{g}$ Kolmogorov-Smirnov probability.
\end{table*}


An obvious way to develop these tests further 
is to compare the predictions of the semi-analytic model
with the MDFs for {\sl both the GCs 
and field-halo stars simultaneously in the same galaxy}.  
A new opportunity to do this has arisen with
NGC 5128, the giant galaxy at the centre of the Centaurus
group at $d \simeq 4$ Mpc \cite{GHarris99}.  
Through several recent HST-based studies, its halo red-giant stars have been 
resolved and used to construct well defined MDFs 
(\citeANP{Soria96} 1996; \citeANP{GHarris99} 1999; 
\citeANP{Marleau00} 2000; \nocite{Harris02} 
\citeANP{Harris00} 2000, 2002).  
In addition, metallicity measurements
based on spectroscopic indices are now available for
$\sim$ 40 of its halo GCs (\citeANP{Held02} 2002).  
The time is thus ripe to combine this observational 
material to test the validity of these models, and see
if a consistent formation history of this galaxy can be 
constructed. 

The paper is organised in the following way:
In Section~\ref{Summary}, we summarise the observational
data used in this study. In Section~\ref{StarFormation}, 
we briefly discuss the aspect of the semi-analytic
model most relevant to this study, namely the star formation
recipe employed. Next, in Section~\ref{Models}, we show example
outputs of the MDFs and star formation histories of eight model
elliptical galaxies and their associated GC systems. In
Section~\ref{Theory}, we pursue a specific comparison between 
one arbitrarily selected galaxy and the observational material.
In Section~\ref{Bursts} we vary the parameter which controls
the threshold for star-bursts in the model to test the robustness of the
results from the fiducial semi-analytic model.
In Section~\ref{AMR}, we propose a possible age-metallicity
relation for the NGC~5128 GCs to be tested against future data
and finally in Section~\ref{Discussion} we summarise our conclusions.

\section{Summary of the Observational Data}
\label{Summary}

For the stellar halo of NGC 5128, MDFs based on large samples of stars
have been obtained through HST/WFPC2 $(V,I)$
photometry.  A finely spaced grid of red-giant evolutionary tracks,
calibrated against Milky Way GCs, is then superimposed on the
colour-magnitude diagram for the NGC 5128 stars and used to define
the stellar metallicity distribution (\citeANP{GHarris99} 1999; 
\citeANP{Harris00} 2000, 2002).
To date, MDFs are available for three different locations in the NGC 5128
halo. The first two are located at projected 
galactocentric distances of 21 and 31 kpc 
(\citeANP{GHarris99} 1999, \citeANP{Harris00} 2000)
and so are likely to be representative of only the
outer halo.  The third of these is a field located at 8 kpc which therefore
samples a combination of the halo and bulge
(see the discussion of \citeANP{Harris02} 2002).  
For our present purposes, we will
use this inner (halo plus bulge) sample  
to define a broadly based MDF representing a plausible average over
the galaxy as a whole.  

In this paper we compare our model results for GCs with the
spectroscopically derived metallicities of \citeANP{Held02} (2002).
These indices are transformed into [m/H] $\equiv$ 
log $(Z/Z_{\odot}$) as described in
\citeANP{Harris02} (2002), yielding $\sim$40 
GCs drawn widely from all over the NGC 5128 halo.  
We note that the GC photometric metallicities from the 
$(C-T_1)$ indices of \citeANP{Harris92} (1992) and 
the $(U-V)$ indices of \citeANP{Rejkuba01} (2001) are consistent
with the \citeANP{Held02} (2002) GC metallicity distribution.
All three of the above GC studies probe regions well beyond 
the central dust lane (which contains a small number 
of $\le$ 1 Gyr old GC candidates associated with a recent merger 
event, \eg see \citeANP{Holland99} 1999).
The [m/H] histograms for both the 
field-star and spectroscopic GC sample are shown in
Figure~\ref{fig1}.

Although small-number statistics may still be affecting our estimates
of the true proportions of metal-poor and metal-rich GCs
(lower panel of Fig~\ref{fig1}), the GC and field-star MDFs are clearly
quite different.  Roughly half the GCs are more metal-poor than
[m/H]= --1, whereas remarkably small numbers of field
stars populate this low-metallicity range.  
Another way to quantify this is to say that the {\sl specific frequency}
of GCs to field stars is a function of metallicity
(\citeANP{Harris02} 2002).
The stellar-halo MDF (upper panel of Fig.~\ref{fig1}) has a 
single major peak at [m/H] $\simeq$ --0.4, though it is 
clearly not symmetric, having a
long and thinly populated low-metallicity tail.  
These two very different MDFs -- stellar and GC system -- must be 
matched by the semi-analytic model simultaneously.

\section{Star Formation in the Model}
\label{StarFormation}

The semi-analytic model we test in this
study is the fiducial version
of \textsc{galform}, as described in detail
by \citeANP{Cole00} (2000)(where $h$=0.7, with $H_0=100~h~\kms
\Mpc^{-1}$, $\Omega_{\rm 0}$=0.3, $\Omega_{\rm b}$=0.7, 
$\Lambda_{\rm 0}$=0.7).
Its application to the modelling of the 
GC systems of elliptical galaxies is presented in 
\citeANP{Beasley02} (2002).

This fiducial version of the \citeANP{Cole00}
model was developed to best match
local galaxy properties, in particular the local $\B$-band
luminosity function. As such, the model should possess genuinely
predictive powers. Therefore, for the majority of this study, 
we retain the identical parameter values 
given in \citeANP{Cole00} (2000).
However, in Section~\ref{Bursts} we vary the parameter which
controls the burst threshold to test the robustness of the form of
the MDF in the model.

Since we are examining the stellar metallicity distributions 
of the model ellipticals here, it is the details of the 
model star formation that we are primarily interested in here.
We briefly discuss the salient points, although for a full
description see \citeANP{Cole00} (2000).
In the model, the PGDs
are modelled as rotationally supported, exponential 
gas discs. This is for convenience only, whereas 
in reality they may be expected to form a continuum 
of objects from flattened discs, to more clumpy systems
dominated by random motions. To avoid 
possible confusion with spiral discs, we continue to refer 
to these objects as ``PGDs''.

Star formation in the model proceeds in two modes, quiescently in
PGDs and during the merging of these PGDs.
The instantaneous star formation rate ($\psi$) 
occurring in these PGDs (the quiescent mode)  
is proportional to the mass of
cold gas available (M$_{\rm cold}$)

\begin{equation}
\label{eq1}
\psi = \frac{M_{\rm cold}}{\tau_*}
\end{equation}
 
\noindent where $\tau_*$ is the star formation
timescale, which is defined as

\begin{equation}
\label{eq2}
\tau_* = \epsilon_*^{-1} \tau_{\rm PGD}\ (V_{\rm PGD}/200 \kms)^{\alpha_*}
\end{equation}

\noindent here, $\tau_{\rm PGD} \equiv r_{\rm PGD} / V_{\rm PGD}$, 
where $V_{\rm PGD}$ and $r_{\rm PGD}$ are the circular velocity
and half-mass radius of the PGDs respectively, whilst
$\epsilon_*^{-1}$ and $\alpha_*$ are dimensionless 
parameters in the model.

Star formation also occurs during major mergers, controlled by
the parameter $f_{\rm ellip}$, which is the mass ratio
of the merger progenitors. In the fiducial model of
\citeANP{Cole00} (2000), $f_{\rm ellip}=0.3$,
corresponding to a minimum mass fraction of 30\% required to
induce a major merger.
During these major mergers, star formation proceeds similarly to 
that occurring quiescently in the PGDs, 
but is based upon the properties 
of the central spheroid (\ie its dynamical time) 
rather than the properties of the merging PGDs. 
In the fiducial model used here $\alpha_*=$0, 
which as discussed in \citeANP{Cole00} (2000), 
yields a $\tau_* \propto \tau_{\rm
PGD}$ scaling in equation~\ref{eq2}, as suggested by 
results such as those of \citeANP{Kennicutt98} (1998).

In lockstep with this star formation, stellar winds and 
SNe are assumed to `feedback' energy
and mass into the IGM, re-heating cooled gas and depositing
metals into the hot gas reservoir of the halo.
The level of mass deposition from these PGDs
is governed by the star formation
rate and the feedback efficiency ($\beta$)

\begin{equation}
\label{eq3}
\rm{d}\it{M_{\rm eject}} = \beta\psi\ \rm{d}\it{t}
\end{equation}

\noindent with this feedback efficiency defined
in terms of the PGD circular velocity

\begin{equation}
\label{eq4}
\beta = (V_{\rm PGD} / V_{\rm hot})^{-\alpha_{\rm hot}}
\end{equation}

\noindent where $\alpha_{\rm hot}$ is a dimensionless
parameter, and $V_{\rm hot}$ is a parameter with 
units of $\kms$.

\begin{figure*}
\centerline{\psfig{figure=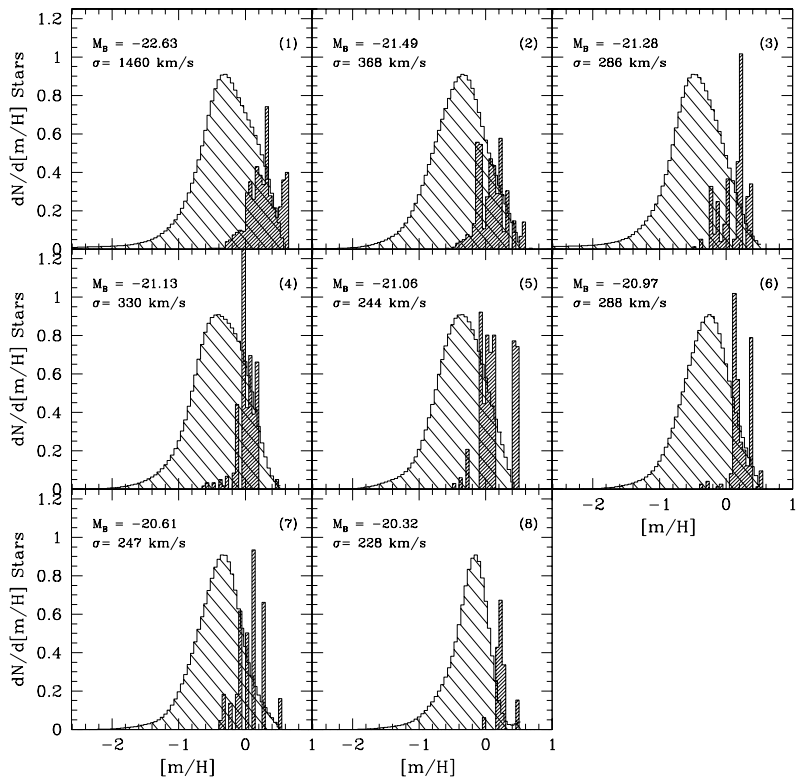,width=4.5in,height=4.5in}}
\caption{\label{fig2} 
Sample metallicity distribution functions for the stars in eight
semi-analytic model galaxies. 
In each panel, the $\B$-band magnitude ($M_B$) of the galaxy
and the velocity dispersion ($\sigma$) of the halo in which it resides 
is labelled at upper left. The numbers in parenthesis 
in the upper identify the number of the model galaxy listed in Table~1. 
The light shaded part of each histogram
shows the metallicity distribution for stars formed in the 
quiescent mode, while the dark shaded histogram shows the stars formed in 
the burst mode (see text). }
\end{figure*}

In the above recipe, up to $\sim$ 30\% of the total stellar mass 
of the model galaxies is composed of burst-formed stars, the
rest originate from the quiescent mode.
This, however, is not the only star formation prescription
used in semi-analytic modelling.  
\citeANP{Somerville99} (1999) present a 
discussion of the principal
star formation recipes used in contemporary
semi-analytic models, and we refer the reader to this
paper for a comparison of the different techniques.

\section{Sample Model Galaxies}
\label{Models}

For this study, we initially make use of the 
450 realisations of model elliptical galaxies 
(bulge-to-total $\geq$ 0.6) described in 
\citeANP{Beasley02} (2002). These model galaxies span a 
$\B$-band magnitude range of --22.7 $\leq$ M$_B$ - 5 log $h$ 
$\leq$ --19.5, and cover a range of environments (halo circular velocity) 
and masses.
For direct comparison with the NGC~5128 MDF, we have selected 
eight model galaxies with luminosities that bracket the 
NGC 5128 luminosity of $M_B \simeq$ --21.0, and look for galaxies 
residing in groups rather than clusters. Beyond these luminosity 
and environmental constraints, the galaxies were chosen arbitrarily.

We list some of the relevant characteristics of NGC~5128 
and the eight selected models in Table~\ref{table1}.
The integrated $B-V$ colours of the model ellipticals 
are very similar to that of NGC~5128\footnote{
It should be noted that due to the large spatial
extent of NGC~5128 on the sky, and the presence of dust in the
disc of this galaxy, the reddening correction is somewhat
uncertain \cite{Israel98}.} ($(B-V)_0$= 0.88; \citeANP{RC3}
1991). The model $V-K$ colours are somewhat redder than those
given in \citeANP{Pahre99} (1999), although these were
determined with unmatched apertures.
This is agreement in the $B-V$ colours is significant, 
since the fiducial \citeANP{Cole00} (2000) model has difficulty
in reproducing the red colours of luminous spheroids
(\citeANP{Cole00} 2000; \citeANP{Benson02} 2002).

\begin{figure*}
\centerline{\psfig{figure=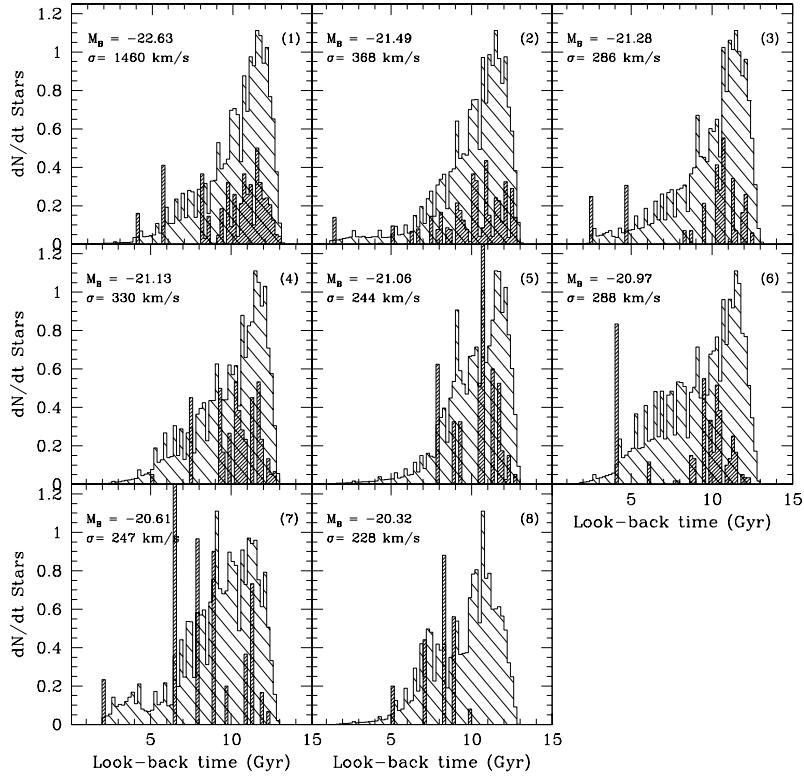,width=4.5in,height=4.5in}}
\caption{\label{fig3} 
Age distributions for the stars in the eight sample models from
Figure 2.  Here, the ages are plotted as
look-back times with the first star formation occurring at $\sim$
13 Gyr. The light filled histograms represent the star formation
in the quiescent mode, the dark filled histograms represent the
contribution from the burst mode.}
\end{figure*}

The output metallicity histograms for the total stellar populations
in eight model galaxies are shown in Figure~\ref{fig2}.
In each case, the distributions reflect the relative number
of stars formed per metallicity bin, for our adopted
yield, feedback efficiency and IMF (in this case Salpeter, 
see \citeANP{Cole00} 2000 for a full list of fiducial 
model parameters).

The most luminous model galaxy (\#1) is a central supergiant in a 
rich cluster, while the two faintest (\#7 and \#8) are 
galaxies smaller than NGC 5128.
These are shown deliberately to help display the range of model outcomes
that are possible.  The other five are model galaxies that are roughly
similar in size to NGC 5128 and drawn from roughly 
Centaurus-like group environments.

The different contributions of the quiescent and burst
modes of star formation are plainly seen in these histograms:  
the ongoing quiescent mode, which takes place within the PGDs before
they merge into the larger proto-galaxy, generate all the 
metal-poor stars in the distribution. It is broad and nearly
featureless, and its range and maxima are very similar from 
one galaxy to the next in our luminosity range.
By contrast, the burst portion which is predominantly 
metal-rich can differ widely between galaxies:  each merger 
produces a $\delta$-function of stars at the metallicity of the
colliding gas, with amplitude in proportion to the amount of 
incoming gas available.  

In Figure~\ref{fig3}, we show the age distributions (look-back times) for
the same eight sample models. Here, it is obvious that both
star formation modes may continue to surprisingly late epochs 
(though not necessarily so; some of the galaxies end up their 
formation quite early, particularly in
the dense environments that give rise to rich galaxy clusters).
Later infall of gas (and subsequent 
bursts) is more prevalent in lower-density
environments.

The galaxy-to-galaxy differences in the burst mode
become particularly obvious for smaller elliptical 
galaxies.  A smaller one
which happened to be built by just one or two 
moderately large mergers will end up with 
a relatively narrow metal-rich stellar component, whilst a galaxy
that was built by a longer sequence of many smaller bursts 
will have a much broader final MDF but will often have a 
rather similar {\sl mean} stellar metallicity.
On the other hand, the biggest ellipticals are almost
always the end result of a large number of individual bursts,
and so the metal-rich portion of the MDF for the biggest ellipticals 
is usually broad and moderately smooth once all these 
bursts are added together.  

Since the quiescently-formed stellar component is fairly similar
amongst all the ellipticals in the semi-analytic
model, it is the merger-formed metal-rich end
which displays most variation from one galaxy to 
another. 
In essence, the metal-rich end of the stellar MDF
should become less broad, more fragmented and with larger 
galaxy-to-galaxy dispersion as we go to lower-luminosity 
galaxies.  

\begin{figure*}
\centerline{\psfig{figure=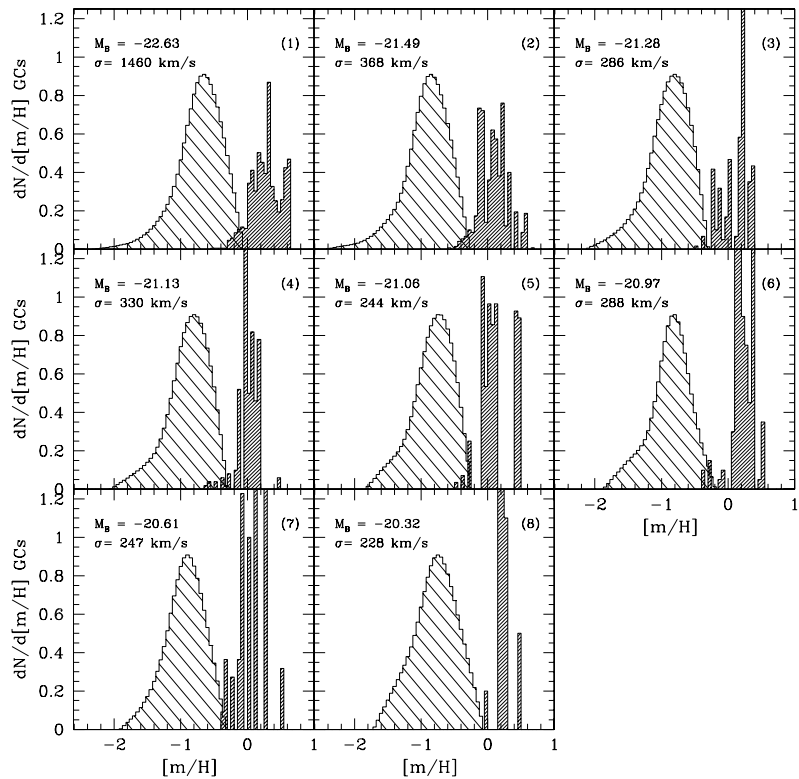,width=4.5in,height=4.5in}}
\caption{\label{fig4} 
Sample metallicity distribution functions for the GCs
in the eight model galaxies plotted in the previous figure.  
The light filled histograms represent the star formation
in the quiescent mode, the dark filled histograms represent the
contribution from the burst mode.}
\end{figure*}

\begin{figure*}
\centerline{\psfig{figure=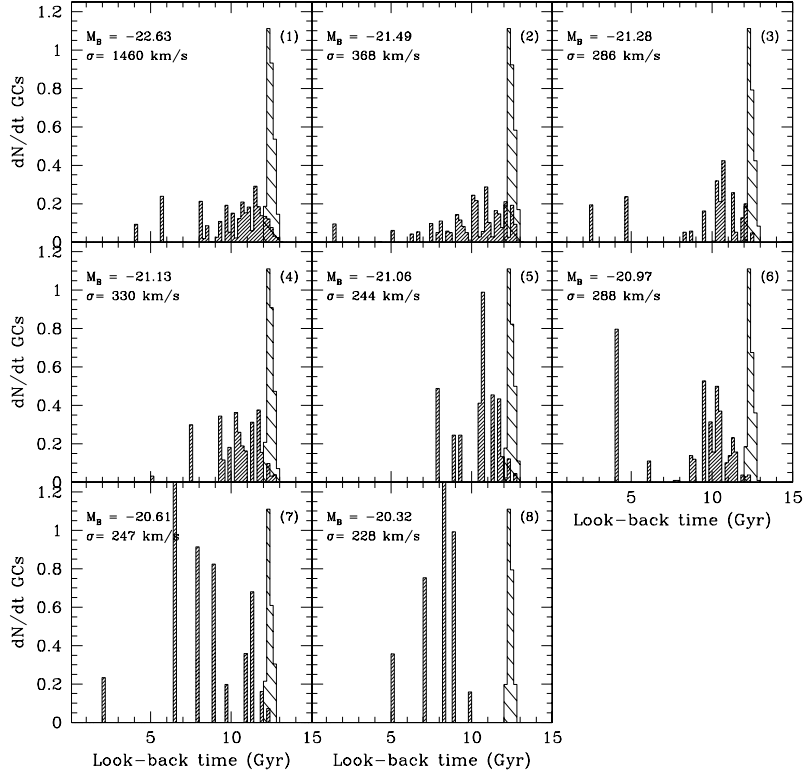,width=4.5in,height=4.5in}}
\caption{\label{fig5} 
Age distributions for the GCs in the eight sample models,
plotted as in the previous figure.
The light filled histograms represent the star formation
in the quiescent mode, the dark filled histograms represent the
contribution from the burst mode.}
\end{figure*}

In Figures~\ref{fig4} and~\ref{fig5} we show the MDFs and age distributions
for the GCs in these same sample models.  In all cases, we
have adopted the parameters $\epsilon_B=0.002$, $\epsilon_R=0.007$,
and $z_{trunc}=5$ as described previously.  Thus for the blue GCs in each
system, the age distribution is very sharply peaked, happening entirely
between 13 Gyr and 12 Gyr look-back time.  It is only for the red GCs that
late formation times occur.  The MDFs are generally 
bimodal in form, although the metal-rich
portions have more complex substructure.

\section{Confronting Theory with Observation}
\label{Theory}

To pursue a direct comparison between the models and
the NGC 5128 data, we superimpose the predicted MDFs 
of the eight model galaxies on these NGC 5128 data.

First, however, we need to account for the fact that 
the full information present in the
model MDF will in observational terms be partly lost by the
smearing effect of measurement uncertainty in [m/H].
The effect is to significantly smooth out the bursty, metal-rich
end of the MDF while doing little to change the broad and already-smooth
metal-poor end.
For the NGC 5128 data from \citeANP{Harris02} (2002), the mean uncertainty is 
$\sigma_{[m/H]}=0.06$ dex.\footnote{The measurement uncertainty in 
heavy-element abundance $Z$, as discussed by Harris \& Harris (2002),
is equivalent to 0.06 dex scatter in [m/H] for the upper giant
branch stars used in their study.  
Since the grid of giant-branch models has a track spacing
of $\sim 0.1$ dex, this uncertainty corresponds to about half a grid step.}

\begin{figure*}
\centerline{\psfig{figure=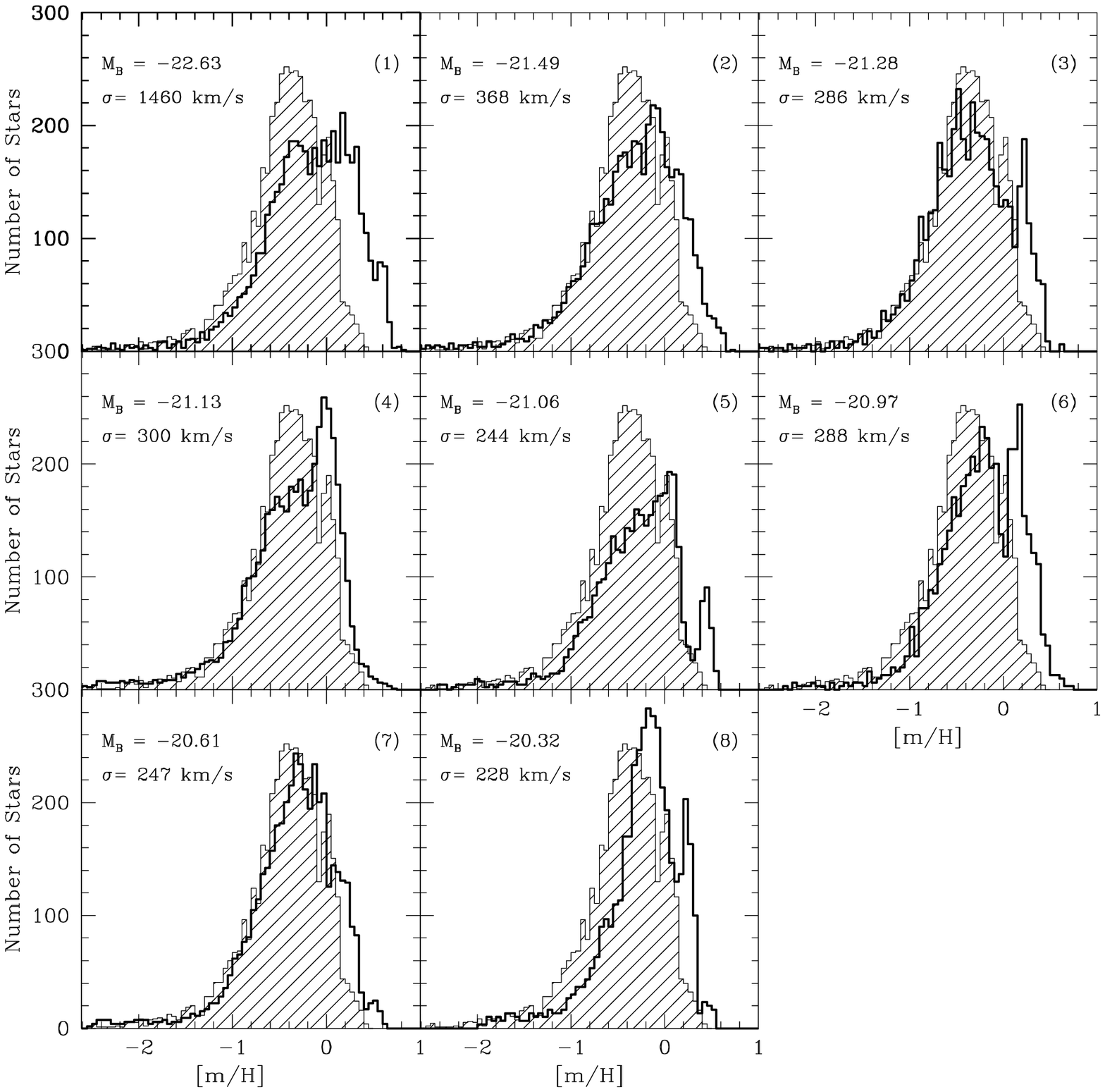,width=4.5in,height=4.5in}}
\caption{\label{fig6} 
The globally averaged model MDFs (solid lines) for the eight selected galaxies
compared to the NGC~5218 data (shaded histograms). 
The models have been normalised to contain the same number 
of stars as these data, and have been smoothed with an error corresponding 
to $\sim$ 0.06 dex.}
\end{figure*}

We show the comparisons between the eight models the and NGC~5128
data in Figure~\ref{fig6}.
It is apparent that in their general characteristics, the
agreement between the MDFs of the models and data 
is reasonably close.
The shape and extent of the low-metallicity tails
are generally similar, with small fractions of stars 
extending  down to [m/H] $\sim$ --2.6.

There is no indication of an over-production 
of metal-poor stars ([m/H] $\leq$ --0.7) 
in the model -- 'the G-dwarf problem' -- 
which is evident when a closed-box model is fit to these 
data (\eg \citeANP{Harris02} 2002).
In contrast to the closed-box `simple model', 
where all heavy elements remain
within the galaxy, the yield, $y$, in the semi-analytic 
model is a function of the level of metal ejection 
into the hot gas phase ($e$) and the feedback efficiency, 
and becomes an effective yield, $y_{\rm eff}$:

\begin{equation}
\label{eq:box1}
y_{\rm eff}  = \frac{(1-e)y}{1-R-\beta}
\end{equation}

\noindent where $R$ is the fraction of mass
recycled by stars. In this way, material processed
by stars is mixed into the 
halo gas reservoir out to large radii.
Therefore, large quantities of metal-poor stars are not 
produced in the semi-analytic model because there is 
continuous infall of gas onto the star forming PGDs
from this cooling reservoir (\citeANP{Kauffmann96a} 1996).

The metal-rich ends of the model MDFs in Figure~\ref{fig6}
also show strong similarities to the NGC~5128 MDF. Both exhibit
significant increases of stars towards high metallicity, 
peaking at or near solar abundances. However, it is at this 
metal-rich end that the differences are also most apparent.
The model MDFs often extend to higher 
metallicities than is the case for the observed MDF. 
This yields mean metallicities between 0.1--0.2 dex
more metal-rich than these data, depending upon the 
model galaxy selected.
In some cases, this excess is manifest as an excess
of stars at $Z>\Zsol$, although for models \#1, \#5 and \#8
in particular it also appears that the effective yield is possibly
too high, since they over-predict the number of stars at 
$\sim$ 0.1$\Zsol$.
Further inspection also reveals that the excess of stars
at the metal-rich ends of the models result in MDFs which
are $\sim$ 0.1 dex broader than these data. 
This is most evident for galaxy \#1, 
the luminous gE galaxy in the centre of a rich cluster.

In assessing these comparisons, it must be born
in mind that the observed MDF almost
certainly underestimates the numbers of high-metallicity stars 
within NGC 5128 {\sl in total} for two reasons:  (a) as is
discussed in Harris \& Harris (2002), the number of red giant
stars with $Z>\Zsol$ found in the 8 kpc field
is probably underestimated because of photometric incompleteness
for these extremely red stars; and (b)
the data samples only one location projected on the outer bulge.

Real elliptical galaxies have radial
abundance gradients (\eg \citeANP{Davies93} 1993;
\citeANP{Gorgas97} 1997), and so 
the inner bulge of NGC 5128 may contain more metal-rich stars.
Both of these effects would raise the upper envelope 
of the observed MDF and increase the metallicity at the MDF peak.  
By contrast, the model galaxies represent 
a globally averaged population because they
contain no spatial information. 
The average abundance gradient of giant ellipticals
is $\sim$ --0.2 $\pm$ 0.1 dex (\eg \citeANP{Davies93} 1993). 
This corresponds to a reduction of 40\% over a factor of 10 in radius,
or a change in metallicity of 0.1 -- 0.3 dex in going from 3 to 30 \kpc. 
However, it remains to be seen whether this gradient 
will be manifest in a global metallicity shift in 
the observed distribution, or result in a change in the
overall shape of the MDF. 

Bearing in mind the dearth of observed stars at $Z>\Zsol$, 
we perform Kolmogorov-Smirnov tests on the eight model galaxies to quantify
their goodness-of-fit with these data.
The resulting probabilities that the models and data are drawn
from the same distribution, $p$(KS), are listed
in Table~\ref{table1}.
In terms of the model, as shown in Figure~\ref{fig2} 
the burst-formed stellar component displays a wide range
of morphologies, with some model realisations providing 
better fits than others.
The best fit is achieved for galaxy \#4, the elliptical 
with $M_B$ - 5 log $h$=--21.13, since the metal-rich 
end of this particular model is fairly narrow compared to
other model galaxies in the sample.


\begin{figure*}
\centerline{\psfig{figure=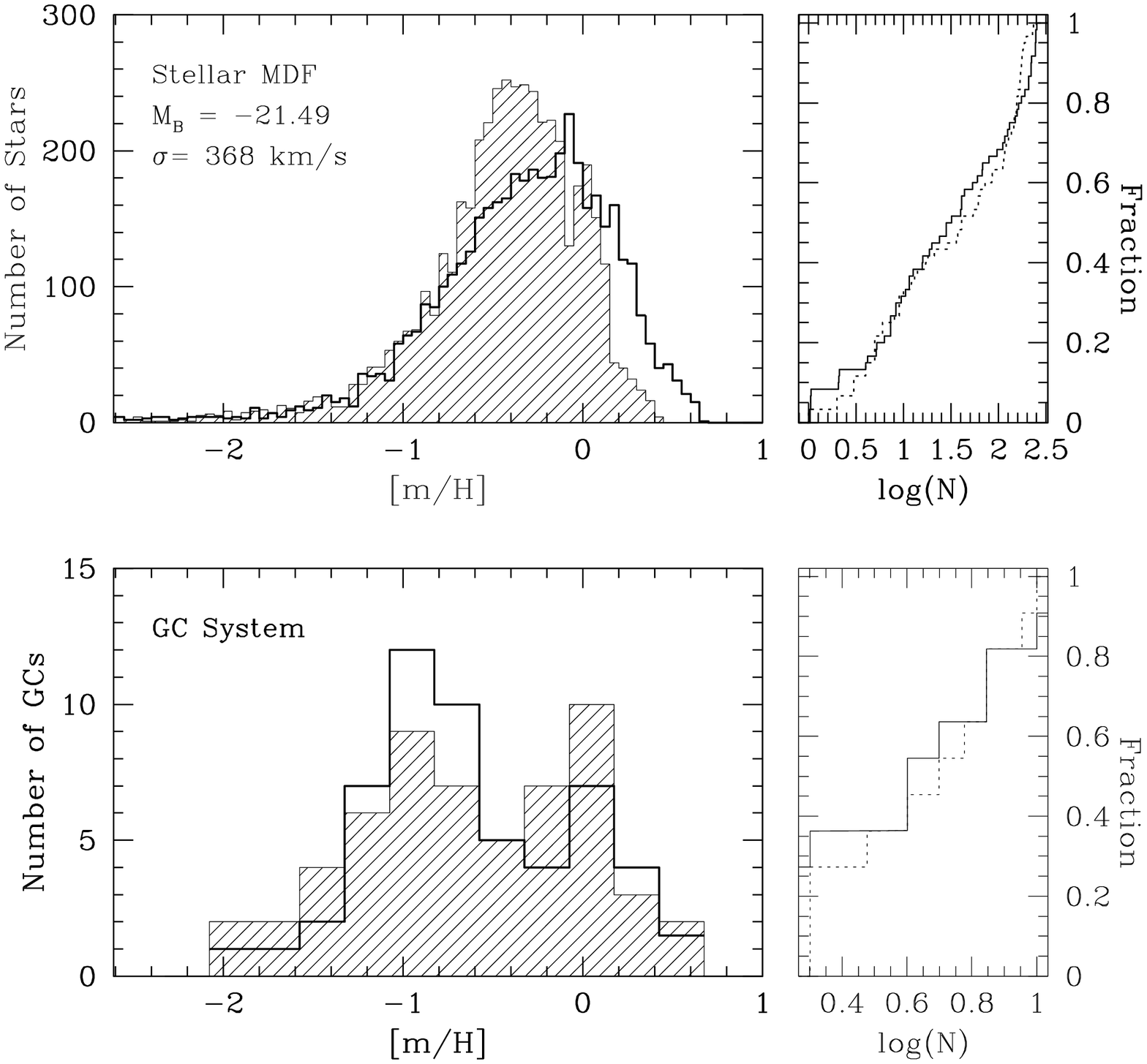,width=5.5in,height=4.5in}}
\caption{\label{fig7} 
{\sl Upper left panel:}  Comparison between a semi-analytic model
stellar metallicity distribution function 
(solid line), and the observed MDF
of old stars in NGC 5128 (shaded histogram).  
The model MDF has been smoothed by the measurement uncertainty
of 0.06 dex as discussed in the text.
{\sl Upper right panel:} Cumulative KS distributions of the model
stars (dotted lines) and data.
{\sl Lower left panel:}  Comparison between a model 
GC MDF (solid line) and the 
observed NGC~5128 GCs (Held \etal 2002; shaded histogram).  
The same smoothing as stated above has been applied to the 
model MDF. 
{\sl Lower right panel:} Cumulative KS distributions of the model
GCs (dotted lines) and data.}
\end{figure*}

We compare a selected model galaxy (in this case
\#2) and its GC system with the NGC~5128 data in Figure~\ref{fig7}.
This galaxy has been chosen somewhat arbitrarily, although
its luminosity and environment is not too dissimilar 
to that of NGC~5128. The excess of metal-rich stars is 
particularly evident in this 
figure. For the GC comparison, we have shifted the 
Held \etal (2002) data by + 0.3 dex in order to approximate a 
conversion between [Fe/H] and [m/H]. 
Both the model and data are clearly bimodal; 
the model peaks at [m/H] $\sim$ --0.95 and --0.05, 
whilst the Held \etal\ dataset has peaks at 
[Fe/H] --1.2 and --0.3, corresponding 
to roughly [m/H] $\sim$ --0.90 and 0.
The metal-rich end of the GC MDF is coincident with that
of the model stellar MDF because both have the same origin
in merger-induced star formation.
As discussed in \citeANP{Beasley02} (2002), a good
fit is achieved at the metal-poor end {\it only} if the
GC formation in PGDs is halted early ($z\sim$5).


\section{The Effect of Star Bursts on the MDF}
\label{Bursts}

Since the primary purpose of this paper is to compare the {\it
predictions} of the semi-analytic model with the observed MDF, 
our previous comparisons have been performed with the
fiducial Cole \etal (2000) model.
However, to test the robustness of these predictions, we 
now change the relative importance of
merger-induced star formation (the burst mode) in the model.
The following is not meant to be an exhaustive exploration of the
parameter space, but rather is a test of the sensitivity of the
model MDF output to the burst assumptions.

Two parameters which have a direct influence on 
the burst mode are $f_{\rm ellip}$ and $f_{\rm burst}$.
As mentioned in Section~\ref{StarFormation}, $f_{\rm ellip}$
represents a mass ratio between the central and satellite
galaxy, which determines whether a merger is classed as 'major'
or 'minor'. Major mergers result in a change in galaxy morphology
(\ie discs are destroyed, stars are redistributed into the bulge)
and in general any cold gas is converted in its entirety into
stars. Minor mergers result in a mass accretion, 
but do not give rise to disc destruction, or generally, star
formation. Any cold gas belonging to the incoming satellite is 
added to the disc of the central galaxy.

The fiducial model of Cole \etal (2000) sets $f_{\rm ellip}$=0.3 (a
30\% central-satellite mass ratio), which is the lower limit
of  the range suggested by numerical simulations of mergers
(\citeANP{Walker96} 1996; \citeANP{Barnes98} 1998).
Increasing $f_{\rm ellip}$ has the principal result of changing
the morphological mix of galaxies in the model; a higher mass
ratio results in fewer major mergers and more 'discy' systems
(\eg \citeANP{Benson02} 2002).
The value of $f_{\rm ellip}$ does not have a significant
effect on the time (or metallicity) of the last merger and 
its star formation, 
since the model ellipticals are generally selected on the basis
that they have undergone a merger recent enough to destroy any
significant disc present.
However, increasing $f_{\rm ellip}$ does decrease the burst-formed
component of the MDFs in the model, which is also
governed by $f_{\rm burst}$.

The parameter $f_{\rm burst}$ sets the {\it threshold} for a 
merger to cause a burst of star formation, and this is  
set to equal $f_{\rm ellip}$ in the fiducial model. 
In this way, a burst of star formation implicitly 
occurs whenever there is a major merger with 
cold gas present.
We can test the effect on the mean galaxian MDFs by changing 
$f_{\rm burst}$ to a smaller value, thus lowering the mass 
threshold required for a burst and increasing the mean
number of bursts per galaxy. 
Setting $f_{\rm burst} < f_{\rm ellip}$ allows
impinging satellites to produce a burst of star formation
without disrupting any disc present.

We have therefore re-simulated $\sim$ 500 early-type galaxies 
(B/T ratio $>$ 0.6) in approximately the same luminosity range 
as the previous realisations 
(--22.7 $\leq$ M$_B$ - 5 log $h$ $\leq$ --19.5).
For this purpose, we use the updated version of the 
semi-analytic code described in \citeANP{Benson02a} (2002).
To facilitate a direct comparison with our previous model 
realisations, we have turned off the photoionizing background contribution 
\cite{Benson02a} .
The updated code produces essentially identical
integrated properties to those of our previous realisations 
for luminous ellipticals at $z=$ 0.

\begin{figure*}
\centerline{\psfig{figure=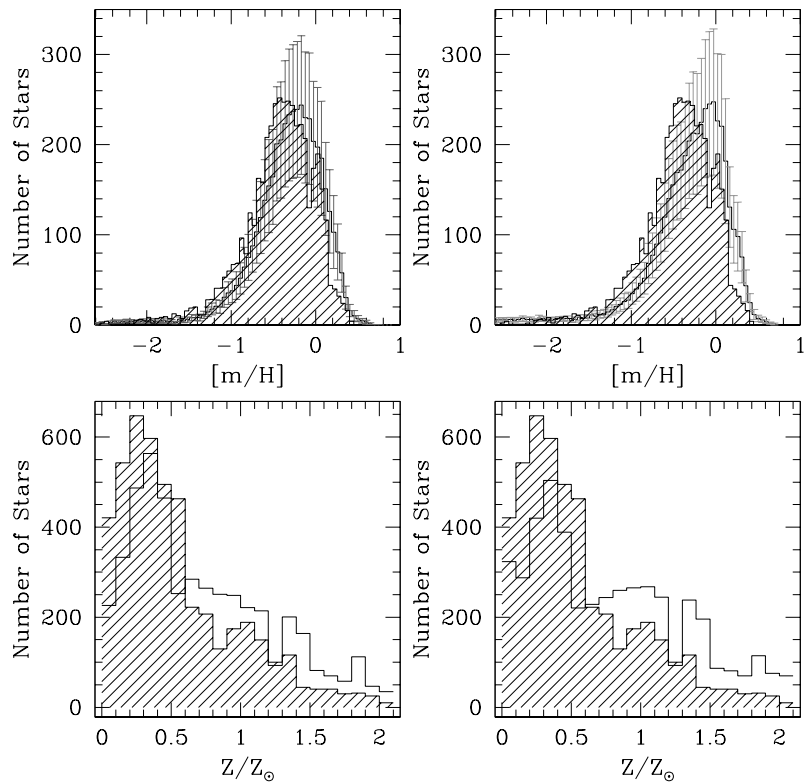,width=4.5in,height=4.5in}}
\caption{\label{fig8} 
{\sl Upper left panel:}  The mean model MDF produced by $\sim$
450 realisations of the fiducial model (with $f_{\rm burst}$=0.3)
compared to the NGC~5128 data (shaded histogram). 
Error bars represent the 1-$\sigma$ scatter on the model
distributions.
{\sl Upper right panel:} The mean model MDF produced by $\sim$
500 realisations of the semi-analytic model with $f_{\rm
burst}$=0.1 (a minimum 10\% central to satellite mass ratio 
is required to induce a star burst) compared to the NGC~5128 data
(shaded histogram). 
{\sl Lower left panel:}  MDFs plotted on a linear scale.
The shaded histogram represents the NGC~5128 data, the
solid line shows the mean model distribution.
{\sl Lower right panel:} Linear version of the upper right plot.}
\end{figure*}

We compare the resulting MDFs with $f_{\rm burst}$=0.3 (left
panels) and $f_{\rm burst}$=0.1 (right panels) in Figure~\ref{fig8}. 
As shown in Section~\ref{Theory}, whilst the form of the fiducial
model MDF is very similar to these NGC~5128 data, the model is $\sim$
0.1 dex offset towards the metal-rich end. The linear plots
(Z/Z$_\odot$) clearly show the excess of higher-metallicity stars.
The model realisations with $f_{\rm burst}$=0.1 (\ie have on
average more bursts of star formation) are slightly more metal
rich again since the stars formed in bursts which contribute to
the model MDF all have around solar metallicities.
However, the differences are not significant; within the
scatter the two sets of model realisations are identical.

In terms of the GCs, the model with $f_{\rm burst}$=0.1 produces
proportionally more metal-rich GCs than the fiducial model used
in Beasley \etal (2002). From comparisons with the well-studied
elliptical NGC~4472, Beasley \etal (2002) found that the
formation efficiency (in the absence of dynamical destruction) of
the metal-rich GCs, $\epsilon_{\rm R}$=0.007.
By applying the normalisation to NGC~4472 described in Beasley
\etal (2002), we find that $\epsilon_{\rm R}$=0.004 for 
$f_{\rm burst}$=0.1. Thus $f_{\rm burst}$ effects the efficiency
of the metal-rich GC formation, $\epsilon_{\rm R}$. 

\section{An Age-Metallicity Relation for NGC~5128?}
\label{AMR}

\begin{figure}
\centerline{\psfig{figure=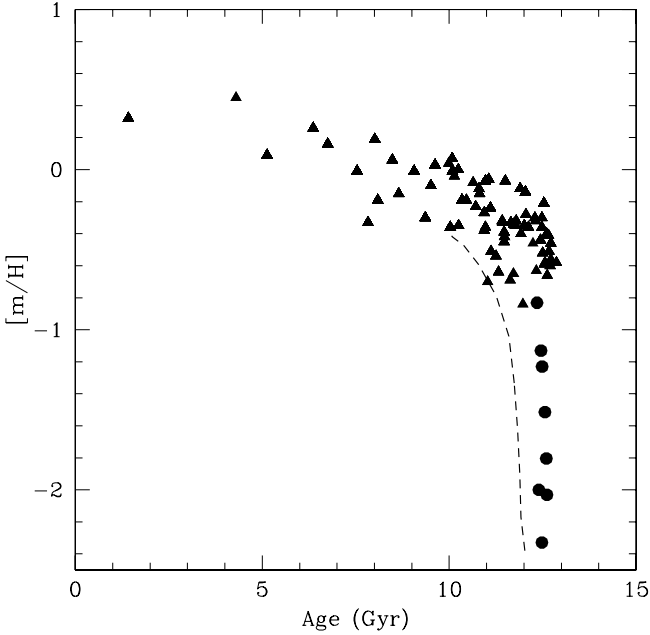,width=3.5in,height=3.5in}}
\caption{\label{fig9} 
The predicted age-metallicity relation (AMR) for our 
NGC~5128-like model galaxy and its GCs. 
The dashed line indicates the AMR for the galaxy, 
which at $z=$0 possesses a mass-weighted mean 
age of 9.5 Gyr, and a metallicity of 0.8 $\Zsol$. 
Filled circles are the (binned) metal-poor GCs, 
which have had their formation truncated at $z=$5.
These have very similar ages ($\sim$ 12.4 Gyr), 
but are able to substantially enrich at early times.
The metal-rich GCs are shown as 
filled triangles, and mark the epoch of each dissipative 
merger undergone by the galaxy. Their formation proceeds to 
late epochs in this galaxy.}
\end{figure}

We show the age-metallicity relation (AMR) of our
model galaxy and its GCs 
in Figure~\ref{fig9}. The dashed
line in the figure indicates the average metallicity
of stars at a given mass-weighted age
in the galaxy. At $z=$0, this galaxy has
a mass-weighted mean age of 9.5 Gyr, and 
a mass-weighted metallicity of [m/H]= --0.1.
Prior to $\sim$ 10 Gyr, the relation for the 
galaxy stars is steep, and rapidly flattens off
at later epochs (\eg \citeANP{Kauffmann96a} 1996). 
This is broadly consistent 
with the behaviour of the solar neighbourhood AMR
(\citeANP{Edvardsson93} 1993; \citeANP{Garnett00} 2000), 
and that of the SMC star clusters \cite{DaCosta98}.

The metal-poor GCs do
not trace the enrichment history of the 
galaxy, but it should be noted that this
feature of the blue clusters is a direct
consequence of the model assumptions (the
formation epoch of these clusters is
deliberately truncated at
$z=$5). However, they clearly are
able to enrich substantially (up 
to [m/H]= --1.0) prior to 
our truncation redshift. This is 
a result of the rapid enrichment 
undergone by 'dwarf' size haloes in the model.

On the other hand, as metal-rich GCs 
are created in each major
merger undergone by the galaxy, each
burst of star formation produces GCs with 
ages and metallicities
corresponding to the epoch of the merger. 
In this scheme, the metal-rich
GCs are discrete
tracers of the dissipational merging
undergone by the galaxy. This is 
in contrast to the luminosity-weighted
properties of the host galaxy since, not only
will age-fading make any burst-formed 
stellar population virtually undetectable 
after a few Gyr (\eg \citeANP{Bruzual01} 2001), 
but also the rapidly decreasing cold gas fraction
in PGDs towards later epochs leads to 
increasingly smaller burst-formed populations
(\eg see figure~2 in \citeANP{Beasley02} 2002).

It is also worth noting in Figure~\ref{fig9}, 
that a fraction of the metal-rich
GCs are slightly more metal-rich
than the {\it global} properties of the galaxy
itself. This occurs because, whilst the
GCs are formed from enriched, well-mixed gas
of a single metallicity, the galaxian stellar 
population comprises of a mixture of 
metal-rich and metal-poor stars 
(\eg Figure~\ref{fig7}). Since merging
is expected to produce new stars
preferentially in the central regions
of the merger product (\eg \citeANP{Mihos96} 1996), 
the central kpc of the galaxy may be
expected to have properties most similar
to the mean properties of the metal-rich
GCs.

Note that the extent of the blue GC AMR is sensitive 
to the truncation redshift we adopt, truncating their 
formation at later epochs results in more metal-rich GCs. 
The red GC AMR is a direct result of the merger histories in 
the hierarchical model.

\section{Discussion and Conclusions}
\label{Discussion}

Semi-analytic models have had some 
success in reproducing the integrated
properties of galaxies over a wide distribution 
in redshift (\eg \citeANP{Kauffmann93} 1993;
\citeANP{Kauffmann96a} 1996; \citeANP{Baugh96} 1996;
\citeANP{Baugh98} 1998).
In this paper, we have pursued a 
rather more specific comparison; that between
the observed and predicted metallicity
distribution functions of the old stars and
GCs of a single elliptical galaxy.

Clearly some discrepancies arose in the
comparison; as noted above, the most obvious
one is that the metal-rich stellar 
component of the model is broader and more metal-rich
than the observed MDF.  
However, it is important to remember we have 
compared an ensemble of model realisations to data for a 
{\it single} galaxy. Whilst integrated colours and spectra give 
some reason to believe the MDFs of ellipticals may be similar,
the MDF of the outer bulge of the Milky Way \cite{Zoccali02}
shows some subtle differences from that of NGC~5128.
Until more star-by-star MDFs are obtained for other ellipticals, 
the existence of significant variations 
(or lack thereof) in their MDFs has yet to be demonstrated.

In terms of the model, we could have altered a number of 
parameters to better fit the galaxy MDF.
For example, decreasing the yield in the model by 0.1 dex 
would act to lower the global metallicity of the model
MDF, better matching the metal-poor tail of the observed 
MDF (and GCs) shown in Figure~\ref{fig7}.
Since the expected theoretical yields for a given
IMF are rather uncertain (\eg \citeANP{Prantzos00} 2000)
this would not be an unreasonable adjustment, although in so
doing may weaken the successes of the model in reproducing
other galaxy properties. 

In any event, obtaining an exact match to the data
was not the purpose of this
study. Rather, our purpose was to see whether the star formation in the 
model leads to an adequate reproduction of the galaxy data, 
and whether in turn we may reconstruct a plausible
star formation history for NGC~5128. 
In this respect, the agreement
has been encouraging. The vast majority of the star
formation in the model occurs 'quiescently', where the 
interplay between gas cooling, star formation and 
feedback yields old stellar populations which are chemically
enriched not unlike those observed in the NGC~5128 bulge/halo. 
Merging of this enriched gas creates an additional metal-rich 
component, including metal-rich GCs (\eg \citeANP{Beasley02} 2002). 

Considering the relative simplicity of the star formation prescription in
the semi-analytic model, the agreement between the observed and
model MDFs is remarkable. We argue that this is also true for the
metal-rich GCs, which fall naturally out of a hierarchical
merging model. Constraining the red GC formation efficiency
through observation is a reasonable undertaking until we have a
consistent theory for globular cluster formation (\eg
\citeANP{Ashman01} 2001).
Formation of the blue GCs in the model 
requires for them to be disconnected from the star formation in 
their progenitors after the truncation redshift 
$z_{\rm trunc}$. Understanding the
formation of the blue GCs will require an understanding of these
high-redshift progenitors (\eg \citeANP{Bromm02} 2002;
\citeANP{Carlberg02} 2002).

To conclude, we find that the metallicity distribution
functions of stars and globular clusters in the nearby 
elliptical galaxy NGC~5128, currently 
the only such galaxy to have its MDF derived on a star-by-star
basis (\citeANP{GHarris99} 1999; \citeANP{Harris00} 2000; 2002), 
are well reproduced by a semi-analytic model of galaxy formation
(\citeANP{Cole00} 2000; \citeANP{Beasley02} 2002).
Since the imprint of relatively old mergers may be hidden in 
stellar populations of galaxies, GCs may provide 
one of the few probes of this star formation.
A large spatially unbiased sample of good quality 
spectra for NGC~5128 GCs will provide an important test of 
this scheme.

\section{Acknowledgements}

We would like to thank Daisuke Kawata, Brad Gibson and 
Chris Brook for lengthy discussions regarding this study, and 
Carlton Baugh for his hard work.
Thanks also to E. Held for sending us his written 
contribution for IAU207 prior to publication, and the anonymous
referee, whose input improved the paper substantially.
This work was supported in part by the Natural Sciences and
Engineering Research Council of Canada, through research grants
to WEH and GLHH, and through Victorian Partnership for Advanced
Computing grants to DAF.

\bibliographystyle{mnras}
\bibliography{mnras} 

\begin{thebibliography}{}

\bibitem[\protect\citeauthoryear{{Arimoto} \& {Yoshii}}{{Arimoto} \&
  {Yoshii}}{1987}]{Arimoto87}
{Arimoto} N.,  {Yoshii} Y., 1987, \aap, 173, 23

\bibitem[\protect\citeauthoryear{{Ashman} \& {Zepf}}{{Ashman} \&
  {Zepf}}{1992}]{Ashman92}
{Ashman} K.~M.,  {Zepf} S.~E., 1992, \apj, 384, 50

\bibitem[\protect\citeauthoryear{{Ashman} \& {Zepf}}{{Ashman} \&
  {Zepf}}{2001}]{Ashman01}
{Ashman} K.~M.,  {Zepf} S.~E., 2001, \aj, 122, 1888

\bibitem[\protect\citeauthoryear{{Barger} et~al.}{{Barger}
  et~al.}{1999}]{Barger99}
{Barger} A.~J., {Cowie} L.~L., {Trentham} N., {Fulton} E., {Hu} E.~M.,
  {Songaila} A.,  {Hall} D., 1999, \aj, 117, 102

\bibitem[\protect\citeauthoryear{{Barnes}}{{Barnes}}{1998}]{Barnes98}
{Barnes} J.~E., 1998, in Saas-Fee Advanced Course 26: Galaxies: Interactions
  and Induced Star Formation, p. 275

\bibitem[\protect\citeauthoryear{{Barnes} \& {Hernquist}}{{Barnes} \&
  {Hernquist}}{1992}]{Barnes92}
{Barnes} J.~E.,  {Hernquist} L., 1992, \araa, 30, 705

\bibitem[\protect\citeauthoryear{{Baugh}, {Cole}, \& {Frenk}}{{Baugh}
  et~al.}{1996}]{Baugh96}
{Baugh} C.~M., {Cole} S.,  {Frenk} C.~S., 1996, \mnras, 283, 1361

\bibitem[\protect\citeauthoryear{{Baugh} et~al.}{{Baugh}
  et~al.}{1998}]{Baugh98}
{Baugh} C.~M., {Cole} S., {Frenk} C.~S.,  {Lacey} C.~G., 1998, \apj, 498, 504

\bibitem[\protect\citeauthoryear{{Beasley} et~al.}{{Beasley}
  et~al.}{2002}]{Beasley02}
{Beasley} M.~A., {Baugh} C.~M., {Forbes} D.~A., {Sharples} R.~M.,  {Frenk}
  C.~S., 2002, \mnras, 333, 383

\bibitem[\protect\citeauthoryear{{Bekki} et~al.}{{Bekki}
  et~al.}{2002}]{Bekki02}
{Bekki} K., {Forbes} D.~A., {Beasley} M.~A.,  {Couch} W.~J., 2002, \mnras, 335,
  1176

\bibitem[\protect\citeauthoryear{{Benson}, {Ellis}, \& {Menanteau}}{{Benson}
  et~al.}{2002}]{Benson02}
{Benson} A.~J., {Ellis} R.~S.,  {Menanteau} F., 2002, \mnras, 336, 564

\bibitem[\protect\citeauthoryear{{Benson} et~al.}{{Benson}
  et~al.}{2002}]{Benson02a}
{Benson} A.~J., {Lacey} C.~G., {Baugh} C.~M., {Cole} S.,  {Frenk} C.~S., 2002,
  \mnras, 333, 156

\bibitem[\protect\citeauthoryear{{Bromm} \& {Clarke}}{{Bromm} \&
  {Clarke}}{2002}]{Bromm02}
{Bromm} V.,  {Clarke} C.~J., 2002, \apjl, 566, L1

\bibitem[\protect\citeauthoryear{{Bruzual} \& {Charlot}}{{Bruzual} \&
  {Charlot}}{2001}]{Bruzual01}
{Bruzual} G.~A.,  {Charlot} S., 2001, in preparation

\bibitem[\protect\citeauthoryear{{C{\^ o}t{\' e}}, {Marzke}, \& {West}}{{C{\^
  o}t{\' e}} et~al.}{1998}]{Cote98}
{C{\^ o}t{\' e}} P., {Marzke} R.~O.,  {West} M.~J., 1998, \apj, 501, 554

\bibitem[\protect\citeauthoryear{{C{\^ o}t{\' e}}, {West}, \& {Marzke}}{{C{\^
  o}t{\' e}} et~al.}{2002}]{Cote02}
{C{\^ o}t{\' e}} P., {West} M.~J.,  {Marzke} R.~O., 2002, \apj, 567, 853

\bibitem[\protect\citeauthoryear{{Carlberg}}{{Carlberg}}{2002}]{Carlberg02}
{Carlberg} R.~G., 2002, \apj, 573, 60

\bibitem[\protect\citeauthoryear{{Cole} et~al.}{{Cole} et~al.}{1994}]{Cole94}
{Cole} S., {Aragon-Salamanca} A., {Frenk} C.~S., {Navarro} J.~F.,  {Zepf}
  S.~E., 1994, \mnras, 271, 781

\bibitem[\protect\citeauthoryear{{Cole} et~al.}{{Cole} et~al.}{2000}]{Cole00}
{Cole} S., {Lacey} C.~G., {Baugh} C.~M.,  {Frenk} C.~S., 2000, \mnras, 319, 168

\bibitem[\protect\citeauthoryear{{Da Costa} \& {Hatzidimitriou}}{{Da Costa} \&
  {Hatzidimitriou}}{1998}]{DaCosta98}
{Da Costa} G.~S.,  {Hatzidimitriou} D., 1998, \aj, 115, 1934

\bibitem[\protect\citeauthoryear{{Davies}, {Sadler}, \& {Peletier}}{{Davies}
  et~al.}{1993}]{Davies93}
{Davies} R.~L., {Sadler} E.~M.,  {Peletier} R.~F., 1993, \mnras, 262, 650

\bibitem[\protect\citeauthoryear{{de Vaucouleurs} et~al.}{{de Vaucouleurs}
  et~al.}{1991}]{RC3}
{de Vaucouleurs} G., {de Vaucouleurs} A., {Corwin} J.~R., {Buta} R.~J.,
  {Paturel} G.,  {Fouque} P., 1991, in Third reference catalogue of bright
  galaxies (1991)

\bibitem[\protect\citeauthoryear{{Dey} et~al.}{{Dey} et~al.}{1998}]{Dey98}
{Dey} A., {Spinrad} H., {Stern} D., {Graham} J.~R.,  {Chaffee} F.~H., 1998,
  \apjl, 498, L93

\bibitem[\protect\citeauthoryear{{Dubinski}}{{Dubinski}}{1998}]{Dubinski98}
{Dubinski} J., 1998, \apj, 502, 141

\bibitem[\protect\citeauthoryear{{Dunlop} et~al.}{{Dunlop}
  et~al.}{1996}]{Dunlop96}
{Dunlop} J., {Peacock} J., {Spinrad} H., {Dey} A., {Jimenez} R., {Stern} D.,
  {Windhorst} R., 1996, \nat, 381, 581

\bibitem[\protect\citeauthoryear{{Edvardsson} et~al.}{{Edvardsson}
  et~al.}{1993}]{Edvardsson93}
{Edvardsson} B., {Andersen} J., {Gustafsson} B., {Lambert} D.~L., {Nissen}
  P.~E.,  {Tomkin} J., 1993, \aap, 275, 101

\bibitem[\protect\citeauthoryear{{Forbes}, {Brodie}, \& {Grillmair}}{{Forbes}
  et~al.}{1997}]{Forbes97}
{Forbes} D.~A., {Brodie} J.~P.,  {Grillmair} C.~J., 1997, \aj, 113, 1652

\bibitem[\protect\citeauthoryear{{Franx} et~al.}{{Franx}
  et~al.}{1997}]{Franx97}
{Franx} M., {Illingworth} G.~D., {Kelson} D.~D., {van Dokkum} P.~G.,  {Tran}
  K., 1997, \apjl, 486, L75

\bibitem[\protect\citeauthoryear{{Garnett} \& {Kobulnicky}}{{Garnett} \&
  {Kobulnicky}}{2000}]{Garnett00}
{Garnett} D.~R.,  {Kobulnicky} H.~A., 2000, \apj, 532, 1192

\bibitem[\protect\citeauthoryear{{Gonz{\'a}lez}}{{Gonz{\'a}lez}}{1993}]{Gonzal%
ez93}
{Gonz{\'a}lez} J.~J., 1993, Ph.D. thesis.\ Univ.\ Calif., Santa Cruz

\bibitem[\protect\citeauthoryear{{Gorgas} et~al.}{{Gorgas}
  et~al.}{1997}]{Gorgas97}
{Gorgas} J., {Pedraz} S., {Guzman} R., {Cardiel} N.,  {Gonzalez} J.~J., 1997,
  \apjl, 481, L19

\bibitem[\protect\citeauthoryear{{Harris} et~al.}{{Harris}
  et~al.}{1992}]{Harris92}
{Harris} G.~L.~H., {Geisler} D., {Harris} H.~C.,  {Hesser} J.~E., 1992, \aj,
  104, 613

\bibitem[\protect\citeauthoryear{{Harris} \& {Harris}}{{Harris} \&
  {Harris}}{2000}]{Harris00}
{Harris} G.~L.~H.,  {Harris} W.~E., 2000, \aj, 120, 2423

\bibitem[\protect\citeauthoryear{{Harris}, {Harris}, \& {Poole}}{{Harris}
  et~al.}{1999}]{GHarris99}
{Harris} G.~L.~H., {Harris} W.~E.,  {Poole} G.~B., 1999, \aj, 117, 855

\bibitem[\protect\citeauthoryear{{Harris}}{{Harris}}{1991}]{Harris91}
{Harris} W.~E., 1991, \araa, 29, 543

\bibitem[\protect\citeauthoryear{{Harris} \& {Harris}}{{Harris} \&
  {Harris}}{2002}]{Harris02}
{Harris} W.~E.,  {Harris} G.~L.~H., 2002, \aj, 123, 3108

\bibitem[\protect\citeauthoryear{{Held} et~al.}{{Held} et~al.}{2002}]{Held02}
{Held} E.~V., {Federici} L., {Cacciari} C.,  {Tesla} V., 2002, \aap, in prep.

\bibitem[\protect\citeauthoryear{{Hibbard} \& {Yun}}{{Hibbard} \&
  {Yun}}{1999}]{Hibbard99}
{Hibbard} J.~E.,  {Yun} M.~S., 1999, \apjl, 522, L93

\bibitem[\protect\citeauthoryear{{Holland}, {C{\^ o}t{\' e}}, \&
  {Hesser}}{{Holland} et~al.}{1999}]{Holland99}
{Holland} S., {C{\^ o}t{\' e}} P.,  {Hesser} J.~E., 1999, \aap, 348, 418

\bibitem[\protect\citeauthoryear{{Hui} et~al.}{{Hui} et~al.}{1995}]{Hui95}
{Hui} X., {Ford} H.~C., {Freeman} K.~C.,  {Dopita} M.~A., 1995, \apj, 449, 592

\bibitem[\protect\citeauthoryear{{Israel}}{{Israel}}{1998}]{Israel98}
{Israel} F.~P., 1998, \aapr, 8, 237

\bibitem[\protect\citeauthoryear{{Jimenez} et~al.}{{Jimenez}
  et~al.}{1999}]{Jimenez99}
{Jimenez} R., {Friaca} A.~C.~S., {Dunlop} J.~S., {Terlevich} R.~J., {Peacock}
  J.~A.,  {Nolan} L.~A., 1999, \mnras, 305, L16

\bibitem[\protect\citeauthoryear{{Kauffmann}}{{Kauffmann}}{1996}]{Kauffmann96a}
{Kauffmann} G., 1996, \mnras, 281, 475

\bibitem[\protect\citeauthoryear{{Kauffmann} \& {Charlot}}{{Kauffmann} \&
  {Charlot}}{1998}]{Kauffmann98a}
{Kauffmann} G.,  {Charlot} S., 1998, \mnras, 297, L23

\bibitem[\protect\citeauthoryear{{Kauffmann}, {White}, \&
  {Guiderdoni}}{{Kauffmann} et~al.}{1993}]{Kauffmann93}
{Kauffmann} G., {White} S.~D.~M.,  {Guiderdoni} B., 1993, \mnras, 264, 201

\bibitem[\protect\citeauthoryear{{Kawata}}{{Kawata}}{2001}]{Kawata01}
{Kawata} D., 2001, \apj, 558, 598

\bibitem[\protect\citeauthoryear{{Kennicutt}}{{Kennicutt}}{1998}]{Kennicutt98}
{Kennicutt} R.~C., 1998, \apj, 498, 541

\bibitem[\protect\citeauthoryear{{Koda}, {Sofue}, \& {Wada}}{{Koda}
  et~al.}{2000}]{Koda00}
{Koda} J., {Sofue} Y.,  {Wada} K., 2000, \apj, 532, 214

\bibitem[\protect\citeauthoryear{{Kuntschner} \& {Davies}}{{Kuntschner} \&
  {Davies}}{1998}]{Harald98}
{Kuntschner} H.,  {Davies} R.~L., 1998, \mnras, 295, L29

\bibitem[\protect\citeauthoryear{{Larsen} et~al.}{{Larsen}
  et~al.}{2001}]{Larsen01}
{Larsen} S.~.~S., {Brodie} J.~P., {Huchra} J.~P., {Forbes} D.~A.,  {Grillmair}
  C.~J., 2001, \aj, 121, 2974

\bibitem[\protect\citeauthoryear{{Lia}, {Portinari}, \& {Carraro}}{{Lia}
  et~al.}{2002}]{Lia02}
{Lia} C., {Portinari} L.,  {Carraro} G., 2002, \mnras, 330, 821

\bibitem[\protect\citeauthoryear{{Marleau} et~al.}{{Marleau}
  et~al.}{2000}]{Marleau00}
{Marleau} F.~R., {Graham} J.~R., {Liu} M.~C.,  {Charlot} S.~., 2000, \aj, 120,
  1779

\bibitem[\protect\citeauthoryear{{Menanteau}, {Abraham}, \&
  {Ellis}}{{Menanteau} et~al.}{2001}]{Menanteau01}
{Menanteau} F., {Abraham} R.~G.,  {Ellis} R.~S., 2001, \mnras, 322, 1

\bibitem[\protect\citeauthoryear{{Mihos} \& {Hernquist}}{{Mihos} \&
  {Hernquist}}{1996}]{Mihos96}
{Mihos} J.~C.,  {Hernquist} L., 1996, \apj, 464, 641

\bibitem[\protect\citeauthoryear{{Naab} \& {Burkert}}{{Naab} \&
  {Burkert}}{2001}]{Naab01}
{Naab} T.,  {Burkert} A., 2001, \apjl, 555, L91

\bibitem[\protect\citeauthoryear{{Nolan} et~al.}{{Nolan}
  et~al.}{2001}]{Nolan01}
{Nolan} L.~A., {Dunlop} J.~S., {Kukula} M.~J., {Hughes} D.~H., {Boroson} T.,
  {Jimenez} R., 2001, \mnras, 323, 308

\bibitem[\protect\citeauthoryear{{Pahre}}{{Pahre}}{1999}]{Pahre99}
{Pahre} M.~A., 1999, \apjs, 124, 127

\bibitem[\protect\citeauthoryear{{Pearce} et~al.}{{Pearce}
  et~al.}{1999}]{Pearce99}
{Pearce} F.~R. et~al., 1999, \apjl, 521, L99

\bibitem[\protect\citeauthoryear{{Prantzos}}{{Prantzos}}{2000}]{Prantzos00}
{Prantzos} N., 2000, New Astronomy Review, 44, 303

\bibitem[\protect\citeauthoryear{{Rejkuba}}{{Rejkuba}}{2001}]{Rejkuba01}
{Rejkuba} M., 2001, \aap, 369, 812

\bibitem[\protect\citeauthoryear{{Richter}, {Sackett}, \& {Sparke}}{{Richter}
  et~al.}{1994}]{Richter94}
{Richter} O.-G., {Sackett} P.~D.,  {Sparke} L.~S., 1994, \aj, 107, 99

\bibitem[\protect\citeauthoryear{{Schweizer}}{{Schweizer}}{1987}]{Schweizer87}
{Schweizer} F., 1987, in Nearly Normal Galaxies. From the Planck Time to the
  Present, p.~18

\bibitem[\protect\citeauthoryear{{Shioya} \& {Bekki}}{{Shioya} \&
  {Bekki}}{1998}]{Shioya98}
{Shioya} Y.,  {Bekki} K., 1998, \apj, 504, 42

\bibitem[\protect\citeauthoryear{{Silk}}{{Silk}}{1977}]{Silk77}
{Silk} J., 1977, \apj, 214, 152

\bibitem[\protect\citeauthoryear{{Somerville} \& {Primack}}{{Somerville} \&
  {Primack}}{1999}]{Somerville99}
{Somerville} R.~S.,  {Primack} J.~R., 1999, \mnras, 310, 1087

\bibitem[\protect\citeauthoryear{{Somerville}, {Primack}, \&
  {Faber}}{{Somerville} et~al.}{2001}]{Somerville01}
{Somerville} R.~S., {Primack} J.~R.,  {Faber} S.~M., 2001, \mnras, 320, 504

\bibitem[\protect\citeauthoryear{{Soria} et~al.}{{Soria}
  et~al.}{1996}]{Soria96}
{Soria} R. et~al., 1996, \apj, 465, 79

\bibitem[\protect\citeauthoryear{{Steinmetz} \& {Navarro}}{{Steinmetz} \&
  {Navarro}}{1999}]{Steinmetz99}
{Steinmetz} M.,  {Navarro} J.~F., 1999, \apj, 513, 555

\bibitem[\protect\citeauthoryear{{Terlevich} \& {Forbes}}{{Terlevich} \&
  {Forbes}}{2002}]{Terlevich02}
{Terlevich} A.~I.,  {Forbes} D.~A., 2002, \mnras, 330, 547

\bibitem[\protect\citeauthoryear{{Thomas}}{{Thomas}}{1999}]{Thomas99}
{Thomas} D., 1999, \mnras, 306, 655

\bibitem[\protect\citeauthoryear{{Trager} et~al.}{{Trager}
  et~al.}{1998}]{Trager98}
{Trager} S.~C., {Worthey} G., {Faber} S.~M., {Burstein} D.,  {Gonzalez} J.~J.,
  1998, \apjs, 116, 1

\bibitem[\protect\citeauthoryear{{van den Bergh}}{{van den
  Bergh}}{2000}]{vandenBergh00}
{van den Bergh} S., 2000, \aj, 119, 609

\bibitem[\protect\citeauthoryear{{Waddington} et~al.}{{Waddington}
  et~al.}{2002}]{Waddington02}
{Waddington} I. et~al., 2002, \mnras, 336, 1342

\bibitem[\protect\citeauthoryear{{Walker}, {Mihos}, \& {Hernquist}}{{Walker}
  et~al.}{1996}]{Walker96}
{Walker} I.~R., {Mihos} J.~C.,  {Hernquist} L., 1996, \apj, 460, 121

\bibitem[\protect\citeauthoryear{{Zepf}}{{Zepf}}{1997}]{Zepf97}
{Zepf} S.~E., 1997, \nat, 390, 377

\bibitem[\protect\citeauthoryear{{Zepf} \& {Ashman}}{{Zepf} \&
  {Ashman}}{1993}]{Zepf93}
{Zepf} S.~E.,  {Ashman} K.~M., 1993, \mnras, 264, 611

\bibitem[\protect\citeauthoryear{{Zoccali} et~al.}{{Zoccali}
  et~al.}{2002}]{Zoccali02}
{Zoccali} M. et~al., 2002, \aap\ in press, (astro-ph/0210660)

\end{thebibliography}

\end{document}